\documentclass[12pt,epsf]{article}
\usepackage{graphicx, amsmath}

\begin{document}

\sloppy
\begin{flushright}{SIT-HEP/TM-34}
\end{flushright}
\vskip 1.5 truecm
\centerline{\large{\bf Generating curvature perturbations}}
\centerline{\large{\bf with or without MSSM flat directions}}   
\vskip .75 truecm
\centerline{\bf Tomohiro Matsuda\footnote{matsuda@sit.ac.jp}}
\vskip .4 truecm
\centerline {\it Laboratory of Physics, Saitama Institute of Technology,}
\centerline {\it Fusaiji, Okabe-machi, Saitama 369-0293, 
Japan}
\vskip 1. truecm
\makeatletter
\@addtoreset{equation}{section}
\def\theequation{\thesection.\arabic{equation}}
\makeatother
\vskip 1. truecm

\begin{abstract}
\hspace*{\parindent}
We consider instant preheating as the mechanism for generating the
 curvature perturbation at the end of chaotic inflation.
Then we examine if inflation could be driven by a
 Minimal Supersymmetric Standard Model (MSSM) flat direction or a
 sneutrino. 
Our simple mechanism relaxes some serious constraints that appeared in
 past studies, making inflation driven by a MSSM flat direction
 or a sneutrino more plausible.
\end{abstract}

\newpage
\section{Introduction}
\hspace*{\parindent}
In the standard scenario of the inflationary Universe, the observed
density perturbations are assumed to be produced by a light inflaton
rolling down its potential during inflation. 
This ``standard mechanism'' of generating the curvature perturbation has
been investigated by many authors \cite{Books-EU}. 
Although it may be possible to construct some inflationary
scenarios with this mechanism, it is not always easy to find a situation
where an inflaton field appears naturally in the Grand Unified Theory
(GUT) of particle physics, and at the same time has all the required
conditions for inflation satisfied without fine-tunings or an additional
hidden sector for inflation. 
In this paper we focus our attention on inflationary models in which a
MSSM flat direction or a sneutrino plays the role of an inflaton. 
As will be presented in sec. \ref{sec:main}, past studies on 
MSSM (or sneutrino) inflation were studied on the basis of a
conventional mechanism of generating the curvature perturbations related
to fluctuations of an inflaton. 
The conditions required for the coupling constants in the theory were
very severe, demanding fine-tunings of the MSSM parameters. 
Hence, our question in this paper is very simple. 
Is it possible to solve or relax the above-mentioned conditions with a
new mechanism for generating curvature 
perturbations? 

Recently, a new inflationary paradigm has been developed where the
conventional slow-roll picture does not play an essential role in
generating the curvature perturbation. 
Along the lines of this new inflationary paradigm, we will consider
a scenario where the perturbation of a light field is converted
into the curvature perturbation at the time of instant preheating that
occurs at the end of inflation, when the inflaton kinetic energy is
significant. 
The most important point is that in this new inflationary paradigm the
light field is not identified with the inflaton. 
The light field is decoupled from the inflationary dynamics during
inflation, but it plays a significant role at reheating. 
The idea related to such a light field has already been investigated by
many authors. 
Although there have been many attempts in this direction, the most
famous examples would be the curvaton models
\cite{curvaton_1,curvaton_2}. 
In the curvaton models, the origin of the large-scale curvature
perturbation in the Universe is the late-decay of a massive scalar field
that is called the ``curvaton''. 
The curvaton paradigm has attracted much
attention because it was thought to have obvious advantages. 
For example, since the curvaton is independent of the inflaton field,
there was the hope \cite{curvaton_liberate} that the curvaton scenario,
especially in models with a low inflationary scale, could cure serious
fine-tunings associated with the inflation models.\footnote{Many
 attempts have been made to construct realistic models with low 
inflationary scale\cite{low_inflation}, where cosmological defects might
 play an essential role\cite{matsuda_defectinfla}.
In spite of the merits of these low-scale inflationary models,
an ultimate solution has not yet been found.} 
Despite the advantages of the curvaton scenario, Lyth suggested
\cite{Lyth_constraint} that there is a strong bound for the Hubble
parameter during inflation, even when the curvatons are introduced to
the model. 
The bound obtained by Lyth \cite{Lyth_constraint} was critical, but it
was later suggested by Matsuda \cite{matsuda_curvaton} that 
the difficulty could be avoided if an additional inflationary expansion
or a phase transition were present \cite{curvaton_added}. 
At first this solution seemed quite generic since there could be several
phase transitions after inflation, but it was later discovered
that this scenario requires many additional components.
It is possible to solve the problem \cite{Lyth_constraint}
 with the idea suggested in Ref.\cite{matsuda_curvaton}, but then one
 cannot escape from uncertainties coming from the additional
 components. 
Besides the curvatons, the problem of uncertainties is common in almost 
all the inflationary scenarios that require fields that cannot appear in
the Standard Model(SM). 
This is one of our motivations to study inflation that occurs on MSSM or
sneutrino direction.\footnote{Our chaotic inflation models can work only
if the value of the inflaton is superplanckian.
Hence in our scenario we expect that the influence of a multitude of
non-renormalizable terms that might appear in the scalar potential 
is substantial.
In our scenario we will consider a tiny Yukawa coupling $y\ll 1$ that can
be related to the very weak violation of an approximate symmetry.
Hence the above expectation would be conceivable if such higher terms
appear with the combination of $\sim |W|^n$ which contains $y^n$, or 
more generically each combination of the corresponding operator
is accompanied by a suppression factor.
However, this clearly pushes the inflationary physics beyond the
justification of the MSSM.}

Another idea in this inflationary paradigm was investigated most
recently by Lyth \cite{delta-N-Lyth}, where an argument was presented
that the density perturbations can be generated at the end of
inflation by the fluctuations of the number of e-foldings induced by
the fluctuations related to a light scalar field other than the
inflaton.\footnote{See also Ref.\cite{alternate}, where 
different models for generating the curvature perturbation
were proposed along the lines of this new inflationary paradigm. 
Although we do not mention these
alternative ideas, these ideas are equally important.}
An important advantage in the new mechanism is that there are no such
stringent conditions coming from the requirement of (1) late-time 
curvaton dominance, and (2) successful reheating. 
Using this new idea, we studied, in
Ref.\cite{matsuda_elliptic,matsuda_ellip2}, the generation of
curvature perturbations in fast-roll inflationary models. 
Since the hybrid-type potential appears naturally in brane-antibrane
inflation and a light field may generically appear in brane inflationary
models, the new mechanism is very useful in brane inflationary models. 
The most obvious example along these lines would be throat inflation,
where a light field related to an enhanced isometry appears at a
distance from the moving brane as we have proposed in
Ref.\cite{matsuda_elliptic}  and \cite{matsuda_ellip2}.\footnote{Riotto
and Lyth made another useful discussions on this
point\cite{Riotto-Lyth}.}
On the other hand, if the brane motion is relativistic, or
the kinetic energy of the inflaton is significant, the so-called
``instant preheating'' should be significant in such analysis. 
In order to solve this problem, we proposed in Ref.\cite{matsuda_PR}
 an alternative mechanism along the lines of this new inflationary
 paradigm. 
This is a generic mechanism for generating the curvature perturbation
that works with an instant preheating at the end of inflaton. 
The instant preheating should be equipped with a light field, whose
expectation value plays the role of an impact parameter. 
This alternative mechanism of generating curvature perturbation at the
end of inflaton will be quite useful in MSSM inflationary models, since
in such models instant preheating is a natural mechanism of reheating
and at the same time there are many flat directions that can play the
role of the light field.   

In this paper we will examine whether the serious conditions that have
been required in past studies can be relaxed by introducing the new
mechanism we presented in Ref.\cite{matsuda_PR}. 
It will be helpful to conduct a short review of previous attempts in
this direction showing why it was difficult to make an inflationary
model on a MSSM (or sneutrino) direction. 
Conclusions are given in sec. \ref{sec:conclusion}. 
We will show that one can solve or relax the serious
conditions that were found in past studies, by introducing this new
mechanism to the theory.

\section{New mechanism and fine-tunings}
\label{sec:main}
\subsection{New mechanism}
Generically a multi-field inflationary model is described by the
background inflaton fields $\phi_{i}(t)$ which evolve according to the
system of coupled differential equations
\begin{equation}
\ddot{\phi_i}+3H \dot{\phi_i}+\frac{\partial V}{\partial \phi_i}=0, 
\, \, \, \, \, \, i=1,...,n
\end{equation}
and 
\begin{equation}
H^2 = \frac{8\pi}{3M_p^2}\left[
\sum_i \frac{\dot{\phi}_i^2}{2} + V \right].
\end{equation}
Without loosing general applicability,
 we can discuss our mechanism with two orthogonal fields, $\phi_1$ and
 $\phi_2$, where $\phi_1$ is a
conventional inflaton field and $\phi_2$ is the light field.
The potential $V(\phi_1, \phi_2)$ is characterized by a hierarchy
between the masses of the fields, and can be modeled by 
\begin{equation}
V(\phi_1, \phi_2)= \frac{m_1^2}{2}\phi_1^2 + \frac{m_2^2}{2}\phi_2^2,
\end{equation}
where $m_1 \simeq O(H)$ and $m_2 \ll m_1$.\footnote{A different approach
has been given by Kolb et.al.\cite{SSB-curvaton}, who assumed 
$m_2 \simeq m_1$}
We consider the instant preheating model\cite{inst_preheating}
as the process through which the inflaton decays into lighter particles.
The typical coupling to the preheat field $\chi$ is written as
\begin{equation}
{\cal L}=\frac{g}{2}(\phi_1^2+\phi_2^2) \chi^2,
\end{equation}
which gives a mass to the preheat field.
Applying the result obtained in Ref.\cite{inst_preheating}, 
the comoving number density $n_{\chi}$ of the preheat field $\chi$
produced during the first half-oscillation of $\phi_1$ becomes
\begin{equation}
\label{n_chi}
n_\chi \simeq \frac{(g|\dot{\phi_1}(t_*)|)^{3/2}}{8\pi^3}
\exp\left[-\frac{\pi g |\phi_2(t_*)|^2}{|\dot{\phi_1}(t_*)|}
\right],
\end{equation}
where $t_*$ is the time when the inflaton $\phi_1$ reaches its minimum
potential at $\phi_1=0$ and where the light field $\phi_2$ may still
have an expectation value $\phi_2(t_*)\ne 0$.
We used $\dot{\phi}_2 =0$ and $\delta
\dot{\phi}_2 =0$ to derive Eq.(\ref{n_chi}).
To obtain an estimate of the curvature perturbation through
Eq.(\ref{n_chi}), we need to write down an expression for $\delta
n_\chi/n_\chi$;
\begin{equation}
\label{nchidelta}
\frac{\delta n_\chi}{n_\chi} = 
\frac{2\pi g |\phi_2(t_*)|^2}{|\dot{\phi}_1(t_*)|} 
\frac{|\delta \phi_2(t_*)|}{|\phi_2(t_*)|},
\end{equation}
where it is assumed that $|\delta\phi_2(t_*)| \ll |\phi_2(t_*)|$ so that
we can neglect higher terms.
To determine the curvature perturbation produced during the decay
process of the preheat field $\chi$, it is sufficient to note that the
generated energy density is proportional to the comoving number density
$n_\chi$. 
Assuming a smooth decay process of the preheat field, the curvature
perturbation $\zeta$ generated during the instant preheating is 
\begin{equation}
\zeta \simeq \alpha \frac{\delta n_\chi}{n_\chi},
\end{equation}
where $\alpha$ is a constant whose numerical value depends on 
how the density of the produced particle redshifts with the Universe
expansion.
Since the field $\phi_2$ is approximately massless during inflation,
the value of the fluctuation is given by $\delta \phi_2 \simeq H_I/2\pi
\simeq V_I^{1/2}/(2\sqrt{3}\pi M_p)$.
In the simplest case of a single-stage inflationary model,
the curvature perturbation $\zeta$ is approximately 
\begin{equation}
\label{xi_1}
\zeta \simeq 
\frac{\alpha 2\pi g |\phi_2(t_*)|^2}{|\dot{\phi}_1(t_*)|} 
\frac{|\delta \phi_2(t_*)|}{|\phi_2(t_*)|}.
\end{equation}
Since the field $\phi_2$ is very light, it is possible to have 
$\phi_2(t_i)\simeq \phi_2(t_*)$, where $t_i$ is the time when the
inflaton $\phi_2$ starts fast-rolling. 
As we are considering a case where the kinetic energy of the inflaton
field $\phi_1$ is significant at the time of preheating, 
\begin{equation}
\label{natu_con}
\dot{\phi}_1(t_*)^2 \simeq m^2 |\phi_1(t_i)|^2 \simeq H_I^2 M_p^2
\end{equation}
is a natural consequence.
We must also consider the condition for the efficient production of the
preheat field $\chi$, which is written as
\begin{equation}
\label{cond_mkin}
m_\chi^2 \simeq g|\phi_2(t_*)|^2 < \dot{\phi}_1(t_*).
\end{equation}
Considering the above conditions, we found a relation
\begin{equation}
\label{curv_fin}
\zeta \simeq \frac{\alpha 2\pi g |\phi_2(t_*)|^2}{|\dot{\phi}_1(t_*)|}
\frac{|\delta \phi_2|}{|\phi_2(t_*)|} 
\simeq \frac{\alpha g |\phi_2(t_*)|}{M_p}
< \frac{\alpha g^{1/2}\sqrt{m \phi_1(t_i)}}{M_p}
\simeq  \alpha g^{1/2}\sqrt{\frac{10 m }{M_p}}.
\end{equation}
Here $\alpha$ depends on the redshift of the final product.
As we are considering instant preheating in a MSSM direction, the final
product is assumed to be massless ($\alpha=1/4$).
The value of $\phi_2$ is not a parameter of the
underlying theory, but an initial condition at the beginning of
inflation.  
The value of $\phi_2$ is important since it determines the value of 
$\zeta$.
The fate of $\phi_2$ is discussed in appendix A.

This is the new mechanism that we will consider in this paper.
With this new mechanism for generating curvature perturbations,
is it possible to construct a chaotic inflation model with or without
MSSM directions and to solve or relax the conditions obtained in past 
studies on MSSM and sneutrino inflation?

\subsection{Quartic potential}
There may be a rather peculiar possibility for chaotic inflation with
MSSM direction, which is to use a D-flat but not an F-flat
direction\cite{moroi-D-flat-MSSM}.
In this scenario we consider quartic potential
\begin{equation}
V(\phi)= \frac{\lambda}{4}\phi^4,
\end{equation}
where $\lambda$ is a dimensionless coupling constant.
The value of the inflaton field when a fluctuation denoted by $k$ exits
horizon is related to the number of e-foldings $N_k$ elapsed after the
horizon exit as\footnote{See appendix A for more details.}
\begin{equation}
\phi_k \simeq \sqrt{8N_k} M_p.
\end{equation}
Hence, using a standard calculation\cite{Books-EU}, the COBE
normalization is given by 
\begin{equation}
\label{stan_la}
\frac{\sqrt{\lambda}}{8}\left(\frac{\phi_k}{M_p}\right)^{3}
\simeq 5.2 \times 10^{-4}.
\end{equation}
which means that normalization of the primordial density fluctuation
requires a tiny value for the coupling constant $\lambda \sim
10^{-13}$\cite{Books-EU}.
One way to solve this problem is to expect an approximate symmetry, such
as the flavor symmetry or R-parity in MSSM.\footnote{R-parity is exact
in MSSM. In this case we are expecting something beyond MSSM.}
If the inflaton field $\phi$ is related to a D-flat direction in MSSM, 
one can obtain the quartic inflaton potential considering the Yukawa
interaction that lifts the D-flat direction. 
Then the coupling constant $\lambda$ is given by $\lambda \simeq y^2$,
where the Yukawa coupling constant is denoted by $y$.
In the present case we have to consider an approximate symmetry that
protects the tiny Yukawa coupling.
Actually, smaller $\lambda$ is more favorable if it is
related to the R-parity violating Yukawa coupling.
We will revisit this issue in appendix A for more details.

We think it is important to notice here that instant preheating would be a
dominant mechanism for reheating in the MSSM inflationary model.
It was discussed by Kasuya et al.\cite{moroi-D-flat-MSSM} that 
instant preheating related to Yukawa couplings is efficient in this
model.
In this case it was suggested that the effect of a A-term is so small
that the inflaton exhibits almost straight-line motion on the complex
plane.
Hence the disturbance of the homogeneous motion caused by the A-term was
neglected. 
In a brane inflationary scenario of ``trapped inflation''\cite{beauty_is},
it has been discussed that open strings stretched between branes play
the role of a preheat field. 
In terms of the brane worldvolume fields the preheat field was 
identified with massive gauge boson that becomes massless at enhanced
symmetry point(ESP).
In the scenario of trapped inflation the W boson was assumed to be a
stable excitation, however in a MSSM inflationary model gauge bosons
 can decay into lighter particles.\footnote{In some other cases the
 amplitude of a oscillation might be much smaller than that in a MSSM
 chaotic inflation, and the oscillation 
might be free from efficient reheating induced by instant preheating.
Typical example would be found in the analysis of Affleck-Dine
baryogenesis\cite{AD}.}
If there is something like a Heisenberg symmetry that protects
flat 
direction from obtaining $O(H)$ mass during inflation, and also if this
direction gives mass to the corresponding preheat field, the impact
parameter for the instant preheating $|\phi_2|$ is non-zero and also it
has a Gaussian fluctuation that exits horizon during inflation.
This is the required situation of our inflationary model.
The curvature perturbation generated with the instant
preheating becomes important in our model.

Despite the novelty of the idea, the actual calculation of the new
mechanism is straightforward. 
The fluctuation of the number density of the preheat field takes
precisely the same formula as Eq.(\ref{nchidelta}), but the kinetic
energy $\dot{\phi}_1$ is different from Eq.(\ref{natu_con}).
Due to the quartic term in the inflaton potential,
$\dot{\phi}_1$ is given by
\begin{equation}
\dot{\phi}_1^2 \simeq \lambda |\phi_1(t_i)|^4 \simeq 10^4 \lambda M_p^4.
\end{equation}
This  relation changes the curvature perturbation obtained
in Eq.(\ref{curv_fin}) as
\begin{equation}
\zeta \simeq \frac{\alpha 2\pi g |\phi_2(t_*)|^2}{|\dot{\phi}_1(t_*)|}
\frac{|\delta \phi_2|}{|\phi_2(t_*)|} 
\simeq \frac{\alpha g |\phi_2(t_*)|}{M_p}
< \alpha g^{1/2}10 \lambda^{1/4}.
\end{equation}
Hence, the curvature perturbation generated at the end
of inflation can be fitted to WMAP data if $\lambda> 10^{-24} g^{-2}$,
which gives the lower bound for $\lambda$.\footnote{Notice that there
are only upper bounds for the couplings related to the R-parity
violating terms, while thre are strict lower bounds for the R-parity
conserving Yukawa terms. See appendix A for more details.}
On the other hand, the ``standard'' curvature perturbation 
(\ref{stan_la}) generated by an inflaton fluctuation becomes large if
$\lambda > 10^{-13}$.
Hence we still need to consider a fine-tuning 
$10^{-24}< \lambda < 10^{-13}$ for $g\sim 1$ to suppress the unwanted
contribution from the inflaton. 
As we have presented above, it is possible to use the new mechanism to
generate the 
curvature perturbation in MSSM inflation.
The obstacle in the above attempt was the large curvature perturbation
generated by an inflaton when $\lambda > 10^{-13}$. 
In the D-flat inflationary model, we found that this problem is solved
if a fine-tuning is introduced to the coupling constant $\lambda$.
Although a kind of fine-tuning is required for the coupling constant
$\lambda$, the experimental upper bounds obtained from phenomenological
considerations are consistent with $10^{-24}< \lambda <
10^{-13}$\cite{moroi-D-flat-MSSM} if $\lambda$ is related
to a R-parity violating Yukawa coupling.
We are expecting that $\lambda$ is protected by an approximate
symmetry, which may (or may not) explain the smallness of $\lambda$. 
As a result, the new mechanism ``relaxes'' the severe condition for
$\lambda$ if it is related to the weak violation of the approximate
R-parity.

\subsection{Quadratic potential}
Perhaps the straightforward extension of the previous example would be
to use a quadratic potential instead of the quartic potential.
The known example in this direction is sneutrino inflation.
Let us first show the known problems and the required fine-tunings in
sneutrino inflation, and then show how one can escape from these
problems using the new mechanism.
Sneutrino is the scalar supersymmetric partner of a heavy singlet
neutrino in the minimal seesaw model of neutrino masses.
The possibility that inflation was driven by a sneutrino has been
discussed by many authors\cite{Sneutrino-inflation}.
Following the past studies, we consider chaotic inflation with a
$V=\frac{1}{2}m^2 \phi^2$ 
potential.
The number of e-foldings $N_e$ is given by
\begin{equation}
N_e\simeq \frac{1}{4}\frac{\phi^2}{M_p^2} \simeq 50,
\end{equation}
which gives the value of $\phi$ at the beginning of inflation,
$\phi(t_i) \simeq {\cal O}(10) M_p$. 
The scale of the inflaton potential is normalized by the WMAP data on
density fluctuations 
\begin{equation}
\label{WMAP-norm}
\zeta \simeq \frac{V^{1/2}}{2\sqrt{6}\pi M_p^2 \epsilon^{1/2}} 
\simeq 10^{-5},
\end{equation}
where 
$\epsilon \equiv \frac{1}{2}M_p^2 \left(\frac{dV/d\phi}{V}\right)^2$
is a slow-roll parameter.
From the above equations we can find the required mass as
$m \simeq 10^{13}$ GeV, which is within the range of heavy singlet
sneutrino masses.
The problem in this scenario is that the reheating temperature 
$T_R \simeq 10^{13}$GeV is much larger than the bound obtained from the
thermal production of gravitinos\cite{gravitino-moroi}.
It may be possible to make $T_R$ much smaller than the bound,
if a neutrino Yukawa coupling $Y_\nu$ is much smaller than unity.
For example, the reheating temperature becomes as low as $T_R\simeq
10^8$GeV if the neutrino Yukawa coupling is $|Y_\nu Y_\nu^{\dagger}|\simeq
10^{-12}$. 
Otherwise, one should expect an additional late-time entropy
production\cite{Books-EU, weak_inf, thermal_inf}, which would be
the secondary inflationary expansion that reduces unwanted
particles.\footnote{Weak inflation\cite{weak_inf} is a mechanism for
such late-time entropy production and is supposed to have the Hubble
parameter not much larger than ${\cal O}$(TeV). There are many ideas for
weak inflation and other kinds of late-time entropy production, however
the most important realization would be the thermal inflation model.
The idea of thermal inflation has been given by Lyth and Stewart in
Ref.\cite{thermal_inf}.}  
However, if there is a late-time entropy production, the baryon number
asymmetry of the Universe must be produced after the entropy production,
which induces another problem in cosmology that may or may not be solved
by introducing additional ingredients to the theory\cite{BAU_after}.

Let us show how this serious condition can be relaxed when
the new mechanism is taken into account.
Our idea is very simple.
%%modified %%
Instead of introducing a tiny Yukawa coupling constant, 
we consider a smaller mass for the sneutrino inflaton.
Then, as we have mentioned above for D-flat inflationary model,
we can assume that reheating is induced by instant preheating. 
Any flat direction that gives a mass for the preheat field 
 can play the role of the impact parameter $\phi_2$, provided that 
the direction is flat.
%%modified %%
Then the dominant contribution comes from Eq.(\ref{curv_fin})
if the inflaton mass is smaller than $10^{13}$GeV.
On the other hand, from Eq.(\ref{curv_fin}) the ratio between $m$ and
$M_p$ is bounded from below, which is given by
\begin{equation}
\frac{m}{M_p} > 10^{-11} g^{-1}.
\end{equation}
This suggests that the inflaton mass $m$ can be as light as
O($10^{7}$GeV) for $g\simeq 1$. 
In the present case the bound for the sneutrino mass is 
$10^{7}$GeV $ < m \le 10^{13}$, which is much looser than the condition
found in the previous study and is suitable to solve the gravitino
problem in sneutrino inflation.

Another important point that we can see from the above analysis is that
the inflaton field $\phi_1$ can now be identified with a 
MSSM flat direction itself.
In this case the inflaton field $\phi$ would be the heaviest direction
which is rather heavier than the lightest direction $\phi_2$
so that inflation ends before the light field $\phi_2$ starts to
roll down the potential.\footnote{When one makes 
predictions in the framework of MSSM, one encounters parameter freedom
which is mainly due to soft SUSY breaking terms. 
The predictive power of the model may be increased if one
restricts this freedom, which is the hypothesis called ``universality of
the soft terms''. 
Under this assumption one is left with 5 free parameters.
However, MSSM with the universal soft masses might not work in
practice. Our inflationary scenario supports non-universal soft
masses with at least one parameter appears at $O(10^7)$GeV.
This is an interesting possibility that requires further study.
See also Ref.\cite{split-susy} in which models with split supersymmetry
have been proposed.
Althernatively, one may consider some extension of MSSM to include a
heavy scalar field.} 
If the dominant part of the curvature perturbation is generated 
at the end of inflation, which occurs if $m < 10^{13}$GeV, 
the scale of the potential is not normalized by Eq.(\ref{WMAP-norm}).

\subsection{A-term}
We would like to make some comments on other inflationary models of MSSM
flat direction.
Recently it has been pointed out that a MSSM flat direction might
support slow-roll inflation with an initial field value much less than
the Planck scale $M_p$ and the tree-level potential
\begin{equation}
V(\phi)= m_\phi^2 \phi^2 +A \cos(n\theta + \theta_0)\frac{\phi^{n+3}}{M^n}
\end{equation}
This potential may have a secondary minimum at 
$\phi_{2nd}=\phi_0 \simeq \left(m_{\phi} M_p^{n-3}\right)^{1/(n-2)} \ll
M_p$, provided that the coefficient of the A-term satisfies the
condition
\begin{equation}
\label{A-term-const}
A \ge A_c \equiv 2\sqrt{2(n-1)}m_{\phi}.
\end{equation}
Moreover, if $A$ takes the critical value $A_c$ 
the first and the second derivatives of $V$ vanishes at $\phi_0$.
If $A = A_c$, the potential near the saddle point is
now very flat along the real direction, and it becomes a successful
inflaton candidate in MSSM potential\cite{MSSM-infla}.
On the other hand, one might think the condition 
$A = A_c$ is a kind of fine-tuning that must be explained by
the underlying (GUT) theory, making the above discussion not within
MSSM. 
We may agree with these critical comments, however the motivation to
make inflationary scenarios on MSSM direction is still very strong in
this scenario.
If the inflaton sector belongs to an unknown (hidden) sector, then there
would be too many uncertainties which really hamper the progress not
only in particle cosmology but also in GUT phenomenology.
On the other hand, if MSSM flat direction could play the role of
inflaton, there will be a chance to prove by some observations 
the existence of such fine-tunings.
If the existence of such fine-tunings may be proved by observations, 
one will be forced to consider a theory behind MSSM that can explain
such fine-tunings. 

Let us examine if our mechanism can help A-term inflation.
Since $V'$ appears in the calculation of 
the number of e-foldings, the fine-tuning related to $V'$ does
always remain.
On the other hand, another condition for the $\epsilon$-parameter
 given by the COBE normalization $V^{1/4}/\epsilon^{1/4}=0.027 M_p$ 
does not appear if the curvature perturbation is generated by an
alternative. 
The most significant condition for the A-term inflationary scenario
comes from the spectral index $n\simeq 1+2\eta$\cite{A-term-cons}
which is related to the $\eta$-parameter, but it does not matter
if the curvature perturbation is generated by an alternative.
As we have stated above, our mechanism requires an additional field
$\phi_2$ that is light ($m_2 \ll H_I$) during inflation.
Since A-term inflation starts with $\phi_0 \ll M_p$, the Hubble
parameter during inflation is $H_I \ll m_{\phi}$.
For example, for $\phi_0 \simeq 10^{-4} M_p$ the potential is 
$V(\phi_0)\sim m_{\phi}^2 \phi_0^2$, and the Hubble constant is
$H_I \sim 10^{-4} m_{\phi}$.
Hence our mechanism requires a light field whose mass satisfies the
condition $m_2 \ll 10^{-4} m_{\phi}$, which introduces another kind of
problem to the model.
Therefore, for successful generation of the
spectrum in A-term inflation one can 
choose either to consider fine-tuning for the shift in the
mass\footnote{See figures in Ref.\cite{A-term-cons}. $m_\phi$ is the
universal soft mass for the components of the inflaton
direction. Fine-tunings for the radiative corrections are included.} 
$\delta^2/m^2_{\phi} \sim 10^{-18}$,
or to introduce a kind of hierarchy $m_2/m_\phi \ll 10^{-4}$.
In the latter case, considering Eq.(\ref{curv_fin}) we found that the
inflaton mass must be as large as $m_\phi>10^{10}$ GeV, and a weak
fine-tuning   is still required for $\delta^2/m^2_{\phi}$ 
to have $N_e\sim 50$.

\section{Conclusions and Discussions}
\hspace*{\parindent}
\label{sec:conclusion}
A new mechanism for generating the curvature perturbation at
 the end of inflation is discussed in this paper.
The dominant contribution to the primordial curvature perturbation
may be generated by this new mechanism, which converts the
perturbation related to a light field into curvature perturbation during
the period of instant preheating.
The light field is ``not'' the inflaton field.
Based on this new inflationary paradigm, 
we considered the possibility that chaotic inflation is driven by a
MSSM flat direction.
We also considered sneutrino inflation as an example for a quadratic
inflationary model.
Our simple mechanism relaxes some serious constraints that appeared in
past studies, making chaotic inflation on a MSSM and a sneutrino
direction more plausible than ever before.

\section{Acknowledgment}
We wish to thank K.Shima for encouragement, and our colleagues at
Tokyo University for their kind hospitality.
We also appreciate the reviewer for many kind instructions that helped
 us to clarify many important points. 
\appendix
\section{Cosmological parameters and MSSM couplings}
\subsection{Basic cosmological parameters}
We consider an arbitrary monomial potential which consists of a single
power of $\phi$;
\begin{equation}
V=\lambda_\alpha \frac{\phi^\alpha}{M_p^{\alpha-4}}.
\end{equation}
Here $\alpha$ is a positive integer.
The slow-roll parameters are
\begin{eqnarray}
\epsilon &=& \frac{\alpha^2}{2}\frac{M_p^2}{\phi^2}\nonumber\\
\eta &=& \alpha(\alpha-1)\frac{M_p^2}{\phi^2}.
\end{eqnarray}
Assuming that inflation ends when $\eta=1$, we obtain
$\phi_e^2=\alpha(\alpha-1)M_p^2$.
Then the value of $\phi$ is given by 
\begin{equation}
\phi_k = \sqrt{\phi_e^2 + 2\alpha N_k} M_p = \sqrt{
\alpha(\alpha-1+2N_k)}M_p \simeq\sqrt{2N_k\alpha}M_p,
\end{equation}
when the cosmological scale related to the number of e-foldings $N_k$
exits horizon.\footnote{Using the exact expression we find $\epsilon = 
\frac{\alpha}{2(\alpha-1+N_k)}$, which suggests that the approximate
expression $\epsilon \simeq \alpha/2N_k$ deviates by a few percent from
the exact value. The exact value should be used for the spectral index.} 

It would be important to note here that 
the chaotic inflation models can work only if the value of the inflaton 
is superplanckian.
This is evident in our present discussions.
At such displacements one expects a multitude of
non-renormalizable terms to appear and become important in the scalar
potential, whose influence on inflation is substantial.
This clearly pushes inflationary physics beyond the justification of the
MSSM.
Therefore, our present model can be applied to the MSSM field contents
while it requires physics beyond the justification of the MSSM.

The COBE normalization is given by
\begin{equation}
\frac{\sqrt{\lambda_\alpha}}{\alpha}\left(\frac{\phi}{M_p}\right)^{\alpha/2+1}
\simeq 5.2 \times 10^{-4}.
\end{equation}
The COBE normalization corresponds to $m \simeq  10^{13}$GeV for the
quadratic potential $V=m^2\phi^2/2$, and to $\lambda \simeq 10^{-13}$
for the quartic potential $V=\lambda \phi^4/4$ \cite{Books-EU}.

For the conventional inflationary scenario the spectral index $n$ is
given by
\begin{equation}
n\simeq 1- 6\epsilon + 2\eta.
\end{equation}
In our present case the spectral index is given by\cite{curvaton_2, 
curvaton_index}
\begin{equation}
n_{PR} \simeq 1 - 2\epsilon_H,
\end{equation}
where the slow-roll parameter $\epsilon_H$ is
\begin{equation}
\epsilon_H \equiv -\frac{\dot{H}}{H^2}.
\end{equation}
As we are considering a very light field $\phi_2$, the corrections from
$\phi_2$ are negligible.
Notice that $n_{PR}$ does depend on the $\epsilon$-parameter of the
inflaton potential, but it does not depend on the $\eta$-parameter
of the inflaton potential.
This is very important for the discussion of
the fine-tuning in the A-term inflation model.
If there is no slow-roll field other than the inflaton, the spectral
index is $n_{PR} \simeq 1-\alpha/2N_k$.
Of course there is an ambiguity in $n_{PR}$ since one cannot simply
disregard the corrections to $\epsilon_H$ from other fields
which might still be rolling down the potential and contribute to
$\epsilon_H$ at the time not so long after the onset of the 
inflationary expansion.

\subsection{MSSM parameters}
We first consider the D-flat direction lifted by the conventional
Yukawa terms that conserve R-parity.
We denote the MSSM superfields as $Q(3,2,\frac{1}{6})$, 
$U(3^*,1,-\frac{2}{3})$, $D(3^*,1,\frac{1}{3})$, $L(1,2,-\frac{1}{2})$,
$E(1,1,1)$, $H_u(1,2,\frac{1}{2})$ and $H_d(1,2,-\frac{1}{2})$,
where the quantum numbers for the $SU(3)_C\times SU(2)_L\times U(1)_Y$
gauge group are shown in the parentheses.
The relevant terms are given by
\begin{equation}
W=(Y_U)_{ij}Q_i U_j H_u + (Y_D)_{ij}Q_i D_j H_d +(Y_E)_{ij}L_i E_j H_d,
\end{equation}
where the indices $i$ and $j$ denotes the generation.
If we consider chaotic inflaton with the quartic potential of
the $Q_i U_j H_u$ direction, the coupling constant $\lambda$ is given by
$\lambda \simeq |(Y_U)_{ij}|^2$.
Since we are expecting small coupling $\lambda \le 10^{-13}$ 
for the quartic inflaton potential, there is the condition $
|Y_{ij}| \le 10^{-7}$ that must be satisfied by the corresponding
Yukawa coupling. 
This condition is so severe that naively the above Yukawa couplings 
cannot satisfy the condition.
However, besides the tree-level contributions,
there could be radiative corrections coming from off-diagonal
(flavor violating) elements.
It has been discussed by Kasuya et. al.\cite{moroi-D-flat-MSSM}
that these corrections may reduce the expected value of the
Yukawa couplings to be within the allowed range of the quartic inflaton
scenario. 

Besides the R-parity conserving couplings that we have shown above,
there could be many kinds of R-parity violating Yukawa
terms.\footnote{See Ref. \cite{R-violating-yukawa} for experimental
bounds.}. 
The experimental bounds suggest that the magnitude of the R-parity
violating Yukawa couplings are typically below $\sim {\cal O}(10^{-7})$.
Hence, the R-parity violating terms are the possible candidates of the 
MSSM inflaton potential if the curvature perturbation is generated by
instant preheating. 

One might think that in the present scenario we expanded the parameter
space of chaotic inflation toward the wrong direction, since this work
allows $\lambda$ to be smaller but not larger than $\lambda\sim
10^{-13}$.
However, there are strict experimental upper bounds on the Yukawa
couplings related to the violation of the approximate R-parity.
Hence the smaller value is more favorable if the inflaton is identified
with such directions.

\subsection{Fate of the light field $\phi_2$}
In the above analysis we assumed that the expectation value of the light
field $\phi_2$ is approximately a constant during inflation.
Due to the exponential factor in $n_\chi$, there is an upper bound for
the expectation value of $\phi_2$,
\begin{equation}
\phi_2 < \sqrt{\dot{\phi}_1} \sim V_I^{1/4}.
\end{equation}
Since in the present case the reheating temperature becomes as large as
$T_R\sim V_I^{1/4}$, the thermal effects from the plasma strongly
affects the dynamics of the flat direction $\phi_2$ evolution after
instant reheating.
Hence in the present case $\phi_2$ evaporates soon after reheating.

\end{document}